\documentclass[USenglish, twocolumn]{article}

\ifx\directlua\undefined\ifx\XeTeXcharclass\undefined
  \usepackage[utf8]{inputenc}                           
  \else\RequirePackage[no-math]{fontspec}[2017/03/31]\fi 
  \else\RequirePackage[no-math]{fontspec}[2017/03/31]\fi 
\usepackage[sort&compress,square,numbers]{natbib}
\usepackage[big,online]{dgruyter}
\usepackage{siunitx}

\theoremstyle{dgthm}

\theoremstyle{dgdef}

\begin{document}

\articletype{Research Article}

\author[1]{Sae R. Endo}
\author[2]{Dasom Kim}
\author[3]{Shuang Liang}
\author[4]{Geon Lee}
\author[5]{Sunghwan Kim}
\author[11]{Alan Covarrubias-Morales}
\author[6]{Minah~Seo}
\author[7]{Michael J. Manfra}
\author*[8]{Dukhyung Lee}
\author[9]{Motoaki Bamba}
\author*[10]{Junichiro Kono} 

\affil[10]{Smalley--Curl Institute, Rice University, Houston, TX 77005, USA; Rice Advanced Materials Institute, Rice University, Houston, TX 77005, USA; Department of Physics and Astronomy, Rice University, Houston, TX 77005, USA; Department of Material Science and NanoEngineering, Rice University, Houston, TX 77005, USA; and Department of Electrical and Computer Engineering, Rice University, Houston, TX 77005, USA, Email: kono@rice.edu. https://orcid.org/0000-0002-4195-0577}
\affil[8]{School of Applied and Engineering Physics, Mohammed VI Polytechnic University (UM6P), Ben Guerir 43150, Morocco, Email: dukhyung.lee@um6p.ma. https://orcid.org/0000-0003-1878-175X}
\affil[1]{Smalley--Curl Institute, Rice University, Houston, TX 77005, USA; and Department of Applied Physics and Physico-Informatics, Keio University, Yokohama 223-8522, Japan. https://orcid.org/0009-0006-9742-5014}
\affil[2]{Department of Electrical and Computer Engineering, Rice University, Houston, TX 77005, USA. https://orcid.org/0000-0002-2331-4526}
\affil[3]{Department of Physics and Astronomy, Purdue University, West Lafayette, IN 47907, USA. https://orcid.org/0000-0002-5801-7521}
\affil[4]{Sensor System Research Center, Korea Institute of Science and Technology (KIST), Seoul 02792, Republic of Korea. https://orcid.org/0000-0002-8569-0260}
\affil[5]{Department of Physics, Ulsan National Institute of Science and Technology (UNIST), Ulsan 44919, Republic of Korea. https://orcid.org/0000-0002-0102-9169}
\affil[11]{Applied Physics Graduate Program, Smalley--Curl Institute, Rice University, Houston, TX 77005, USA; and Department of Electrical and Computer Engineering, Rice University, Houston, TX 77005, USA}
\affil[6]{Department of Physics, Sogang University, Seoul, 04107, Republic of Korea. https://orcid.org/0000-0003-1290-9716}
\affil[7]{Department of Physics and Astronomy, Purdue University, West Lafayette, IN 47907, USA. https://orcid.org/0000-0002-2049-8438}
\affil[9]{Department of Physics, Graduate School of Engineering Science, Yokohama National University, Yokohama 240-8501, Japan; Institute for Multidisciplinary Sciences, Yokohama National University, Yokohama 240-8501, Japan. https://orcid.org/0000-0001-9811-0416}

\title{Cavity-Mediated Coupling between Local and Nonlocal Modes in Landau Polaritons}
\makeatletter
\def\@runningauthor{S.\ R.\ Endo et al.}
\makeatother
\runningtitle{Cavity-Mediated Coupling between Local and Nonlocal Modes in Landau Polaritons}
\abstract{The multimode ultrastrong coupling (USC) regime has emerged as a novel platform for accessing previously inaccessible phenomena in cavity quantum electrodynamics. Of particular interest are cavity-mediated correlations between local and nonlocal excitations, or equivalently, between modes at zero and finite in-plane momentum modes, which offer new opportunities for controlling light--matter interactions across space. However, direct experimental evidence of such interactions has remained elusive. Here, we demonstrate nonlocal multimode coupling in a Landau polariton system, where cavity photons simultaneously interact with the zero-momentum cyclotron resonance and finite-momentum magnetoplasmons of a two-dimensional electron gas in a GaAs quantum well. Our slot cavities, with their subwavelength mode volumes, supply in-plane momentum components that enable the excitation of finite-momentum matter modes. Terahertz time-domain magnetospectroscopy measurements reveal a clear splitting of the upper-polariton branch, arising from hybridization between magnetoplasmon modes and the cavity--cyclotron-resonance hybrids. Extracted coupling strengths confirm USC of the cyclotron resonance and strong coupling of the magnetoplasmon modes to the cavity field, respectively. The experimental results are well captured by the multimode Hopfield model and finite-element simulations. These findings establish a pathway for engineering multimode light--matter interactions involving zero- and finite-momentum matter modes in the USC regime. 

}
\journalname{Nanophotonics}

\maketitle
    Vacuum--matter interactions in cavities are currently explored in diverse systems both for controlling material properties without any external driving field and for applications in quantum information technologies. The ultrastrong coupling (USC) of vacuum--matter arises when the vacuum Rabi frequency becomes a significant fraction of the bare frequencies of vacuum and matter at zero detuning ($\omega_0$), i.e., $g/\omega_0 \geq 0.1$, where $g$ is the coupling strength~\cite{forn-diazUltrastrongCouplingRegimes2019,friskkockumUltrastrongCouplingLight2019}. This regime gives rise to exotic phenomena, via the non-negligible contribution of the counter-rotating terms in the Hamiltonian, such as quantum squeezing in the ground state~\cite{ciutiQuantumVacuumProperties2005, deliberatoVirtualPhotonsGround2017}, the vacuum Bloch--Siegert shift~\cite{liVacuumBlochSiegert2018}, the Dicke superradiant phase transition~\cite{heppSuperradiantPhaseTransition1973, Kim2025}, cavity-mediated superconductivity~\cite{schlawinCavityMediatedElectronPhotonSuperconductivity2019}, and modifications of the quantum Hall effect~\cite{appuglieseBreakdownTopologicalProtection2022, enknerTunableVacuumfieldControl2025a}. The USC regime has been realized in Landau polaritons~\cite{hagenmullerUltrastrongCouplingCavity2010, scalariUltrastrongCouplingCyclotron2012, maissenUltrastrongCouplingField2014, zhangCollectiveNonperturbativeCoupling2016, bayerTerahertzLightMatter2017, kellerFewElectronUltrastrongLightMatter2017, liVacuumBlochSiegert2018, paravicini-baglianiMagnetotransportControlledLandau2019, kellerLandauPolaritonsHighly2020, kuroyamaCoherentInteractionFewElectron2024, mornhinwegSculptingUltrastrongLight2024, scalariUltrastrongLightmatterCoupling2025}, phonon polaritons~\cite{Kim2020,Yoo2021,Roh2023,Kim2024Phonon}, and magnon polaritons~\cite{Li2018,Makihara2021,Kim2025}. Among these systems, Landau polaritons offer exceptional tunability through externally applied magnetic fields. 

Recently, attention has expanded toward the multimode USC regime, where a matter excitation ultrastrongly couples simultaneously with multiple cavity modes, or conversely, a single cavity mode interacts with multiple matter excitations. This regime is not only a natural extension of single-mode USC but also introduces fundamentally new physics due to the nontrivial interplay between cavity modes or between matter excitations~\cite{Balasubrahmaniyam2021,Cortese2023,mornhinwegModemultiplexingDeepstrongLightmatter2024,mornhinwegSculptingUltrastrongLight2024,Kim2024Phonon, Tay2025}. For example, correlations between bare excitations emerge due to the significant contribution of the counter-rotating terms, which do not exist in the absence of ultrastrong light--matter interactions. Very recently, matter--mediated photon--photon correlations~\cite{Tay2025} and cavity-mediated superthermal phonon--phonon correlations~\cite{Kim2024Phonon} have been realized by leveraging the large dipole moment of Landau-level transitions and the subwavelength light confinement of metamaterial cavities, respectively.

Furthermore, nonlocal, i.e., $k \ne 0$ aspects of USC are also drawing increasing interest~\cite{rajabaliPolaritonicNonlocalityLight2021}, where $k$ is the in-plane wave vector of matter excitation. In particular, the possibility of mediating long-range nonlocal correlations through the vacuum field opens a new direction in controlling collective matter excitations across space. Therefore, it would be particularly interesting to explore hybrid quantum states that bridge local and nonlocal matter excitations---coupling between $k=0$ and finite $k$ modes mediated by a cavity mode. 
While recent studies have explored multimode USC involving finite $k$ excitations~\cite{paravicini-baglianiGateMagneticField2017, rajabaliPolaritonicNonlocalityLight2021, kuroyamaCoherentInteractionFewElectron2024, mornhinwegModemultiplexingDeepstrongLightmatter2024}, direct observation of mode-resolved anticrossings between individual modes has so far remained elusive. To further deepen our understanding of multimode USC, it is desirable to investigate systems where such interactions can be more clearly resolved.

Here, we report spectroscopic evidence for coupling between a local mode and a nonlocal mode, through simultaneous USC and strong coupling (SC) with a single cavity mode, in a Landau polariton system. We fabricated an array of slot cavities on a two-dimensional electron gas (2DEG), and its microstructure allowed cavity photons to couple with finite-$k$ plasmon modes of the 2DEG by providing in-plane momentum. In a magnetic field, we experimentally observed SC of magnetoplasmon (MP) modes and the cavity mode, as well as USC between the cyclotron resonance (CR) of the 2DEG and the cavity mode. This multimode hybridization resulted in a clear splitting of the upper-polariton (UP) branch, which can be described by the multimode Hopfield model. Simulated transmission spectra using the finite-element method exhibited similar behavior, confirming the effect of finite-$k$ MP modes. Our findings highlight a new platform where local and finite-$k$ matter excitations can be coupled through a cavity mode, offering a perspective on momentum-resolved multimode interactions in the USC regime.
\begin{figure*}[t!]
\includegraphics[width=\textwidth]{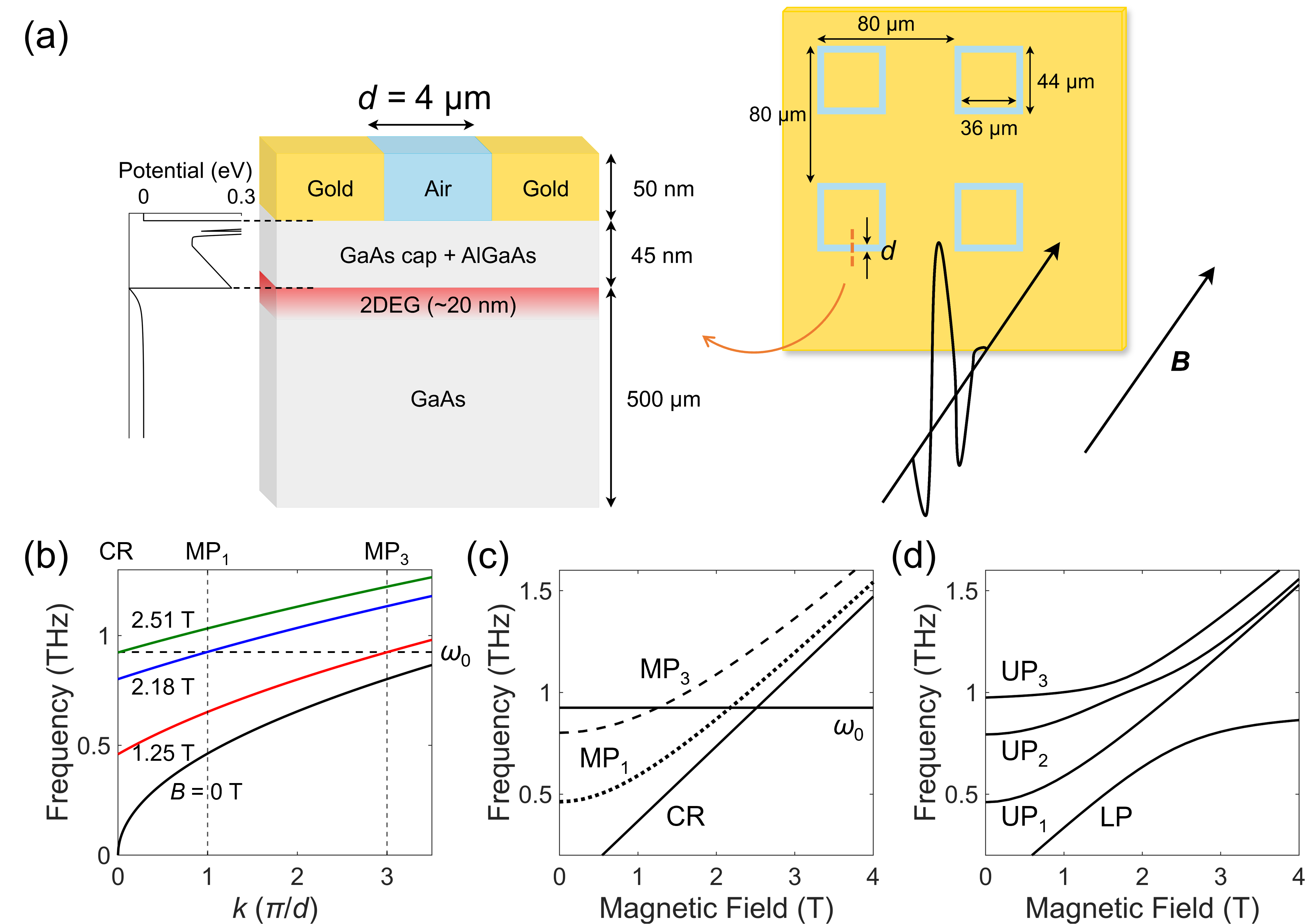} 
\caption{A multimode coupled system involving local and nonlocal matter excitations, and a cavity mode. (a)~Schematic of the GaAs 2DEG--slot system. 2DEG: two-dimensional electron gas. (b)~Dispersion relations of plasmons at 0\,T (black) and~magnetoplasmons at 1.25\,T (red), 2.18\,T (blue), and 2.51\,T (green), corresponding to zero-detuning magnetic fields of cavity mode and MP$_1$, MP$_3$, and CR to $\omega_\text{0}$. $\omega_\text{0}$: bare cavity frequency. (c), (d)~Illustration of the multimode coupling of CR, MP$_1$, and MP$_3$ to the cavity mode. (c)~Uncoupled case. (d)~Coupled case. The multimode coupling leads to the formation of one lower polariton (LP) and three upper polaritons (UP$_1$, UP$_2$, and UP$_3$).} 
\label{fig1}
\end{figure*}

Providing in-plane momentum for light--matter interactions has long been achieved using periodic lattice structures~\cite{allenObservationTwoDimensionalPlasmon1977, theisTwodimensionalMagnetoplasmonSilicon1977, andoElectronicPropertiesTwodimensional1982, batkeNonlocalityTwoDimensionalPlasmon1985, batkePlasmonMagnetoplasmonExcitation1986a, hollandQuantizedDispersionTwoDimensional2004a}, via patterning 2DEGs~\cite{vasiliadouCollectiveResponseMicrowave1993, mikhailovMicrowaveResponseTwodimensional2005, kukushkinCollectiveExcitationsTwodimensional2006, mikhailovInfluenceContactsMicrowave2006, muravevUltrastrongCouplingHighfrequency2013, paravicini-baglianiGateMagneticField2017, paravicini-baglianiMagnetotransportControlledLandau2019, parkTerahertzMagnetoplasmonResonances2021, kuroyamaCoherentInteractionFewElectron2024} or by confining light in a small mode volume~\cite{muravevObservationHybridPlasmonphoton2011, rajabaliPolaritonicNonlocalityLight2021}. To implement this mechanism, we fabricated an array of slot cavities on a GaAs 2DEG by using standard photolithography and lift-off techniques. The 2DEG was present about 45\,nm below the slot, ensuring strong overlap with the electromagnetic fields of the cavity mode; see Fig.\,1(a). The electron density and mobility were $n_\text{e}=3.6 \times 10^{11}$\,cm$^{-2}$ and $\mu_\text{e} = 1.2 \times 10^6$\,cm$^2$/(V$\cdot$s), respectively, obtanined through van der Pauw measurements. The electron cyclotron mass was found to be $m^* = 0.076m_0$ through terahertz (THz) magnetospectro
opy measurements, where $m_0$ is the free electron mass in vacuum. The loop length of slots was tuned to have a resonance frequency in the THz frequency range. The bare cavity frequency, $\omega_\text{0}$, was at 0.925\,THz. The width of the slot, $d$, was 4\,$\upmu$m to confine light inside the cavity, providing cavity photons with finite in-plane momentum. Although the periodic structure also provided finite in-plane momentum, its frequency was too low compared with the cavity mode. Therefore, we consider only the $k$ components generated by the individual slots. 

As a finite-$k$ matter excitation, we had the plasma oscillations of the 2DEG. The 2D plasma frequency in the long-wavelength limit is given by~\cite{sternPolarizabilityTwoDimensionalElectron1967}
\begin{align}
    \omega_\text{p}(k)=\sqrt{\frac{ke^2n_\text{e}}{2m^*\varepsilon_0\varepsilon_\text{r}}}, \label{plasmon}
\end{align}
where $\varepsilon_0$, $\varepsilon_\text{r}$, and $e$ are the vacuum permittivity, the effective dielectric constant of the surrounding medium ($\varepsilon_\text{r} = 6.98$), and the elementary charge, respectively; see Fig.\,1(b) (black solid line). The slot can provide the in-plane momentum $k=n\pi/d$, where $n$ is restricted to odd integers due to the symmetry of the electric field~\cite{mikhailovMicrowaveResponseTwodimensional2005, kuroyamaCoherentInteractionFewElectron2024}. While high-order terms are possible, they are typically weaker and more difficult to observe. Based on our experimental results, we focus on the lowest two components, corresponding to the $n=1$ and $3$ modes. However, even these modes alone do not provide sufficient momentum to match the plasma resonance to the cavity frequency $\omega_0$, resulting in finite detuning, i.e., there is no intersection between the cavity mode and the $n=1$ and 3 MP modes at $k \leq 3\pi/d$.

\begin{figure*}[t!]
\includegraphics[width=\textwidth]{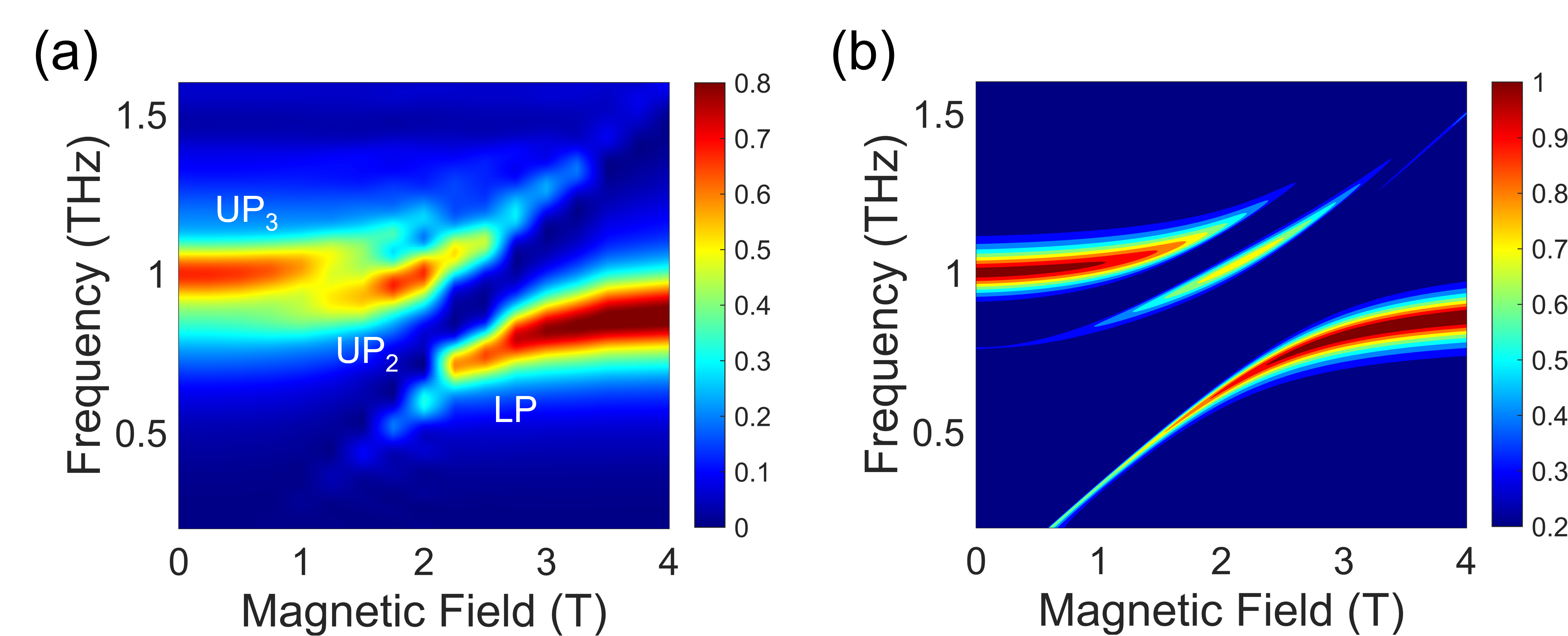}
\caption{Spectroscopic evidence for the multimode coupling. (a)~Experimental color map of transmittance as a function of frequency and magnetic field. The upper polariton is split into two parts, corresponding to UP$_2$ and UP$_3$ in Fig.\,1(d). (b)~Theoretical color map of transmittance based on the multimode Hopfield model.}
\label{fig2}
\end{figure*}

Another matter excitation in our system was the CR of the 2DEG, arising from transitions between adjacent Landau levels at $k=0$, which can be tuned by a perpendicular magnetic field, $B$, through the cyclotron frequency $\omega_\text{c}=eB/m^* \label{cr}$. In the presence of $B$, the 2D plasma excitations evolve into MPs, whose dispersion is given by~\cite{chaplikPossibleCrystallization1972, chiuPlasmaOscillationsTwodimensional1974}
\begin{align}
    \omega_{\text{MP}}^2(k) = \omega_\text{p}^2(k) + \omega_\text{c}^2. \label{magnetoplasmon}
\end{align} 
As shown in Fig.\,1(b) (colored lines), the entire dispersion curve blue-shifts as the magnetic field increases. The colored curves represent the MP dispersion at various magnetic field strengths, with the red (at 1.25\,T), blue (at 2.18\,T), and green lines (at 2.51\,T) indicating the zero-detuning conditions to the cavity mode. These field-tuned resonances can satisfy the momentum matching condition for $n=3$ (MP$_3$), $n=1$ (MP$_1$), and CR, respectively, enabling the exploration of hybridization of $k=0$ and $k\neq0$ matter excitations through the cavity mode.

Figure\,1(c) plots the bare matter excitation frequencies and the cavity frequency as a function of $B$. In this uncoupled situation, the CR and MP modes cross the cavity mode, showing three zero-detuning points, as expected from Fig.\,1(b). However, when the CR and MP modes are coupled with the cavity mode, they are expected to exhibit an anticrossing behavior at each zero-detuning point; see Fig.\,1(d). The multimode coupling gives rise to one lower polariton (LP) and three upper polariton branches (UP$_1$, UP$_2$, and UP$_3$). 

To confirm our model, we performed THz time-domain magnetospectroscopy measurements on the coupled system. THz pulses were generated and detected using InGaAs photoconductive antennas fiber-coupled to an Er-doped fiber laser with a center wavelength of 1550\,nm and repetition rate of 80\,MHz. The electric field strength was sufficiently low to prevent any nonlinear effects. Measurements were conducted at 1.6\,K in a magneto-optical cryostat operating in the Faraday geometry. Time-domain signals of the samples and reference (bare GaAs 2DEG) were Fourier transformed to obtain transmission spectra from 0.2\,THz to 1.6\,THz. 
Figure\,2(a) shows a color map of transmittance as a function of frequency and magnetic field. At around 2.50\,T, an anticrossing behavior is observed, arising from the USC between the cavity mode and the CR of the 2DEG at $k=0$. At around 1.25\,T, an additional splitting of the UP branch is clearly observed. This is evidence of SC of the MP$_3$ mode with the cavity mode having $k=3\pi/d$, induced by the confinement in the 4\,$\upmu$m slot. The split upper polaritons can be interpreted as UP$_2$ and UP$_3$ in Fig.\,1(d). UP$_1$ is absent in the experimental data due to the THz spectra reflecting the cavity-like components of the polaritons, as well as the proximity of UP$_1$ to the CR at finite fields. 

To theoretically explain the coupling between a cavity mode, CR, and MP modes, we introduce a multimode Hopfield model. By adding the MP modes to the full Hamiltonian of Landau polaritons presented in~\cite{liVacuumBlochSiegert2018} while considering only a single cavity mode, the multimode Hamiltonian can be written as 
\begin{align}
    \frac{\widehat{H}}{\hbar} =& \sum_{\xi=\pm}{\omega_{0}{\hat{a}}_\xi^\dagger\hat{a}_\xi} + \omega_{\text{c}}\hat{b}^\dagger\hat{b} + \sum_{n}{\omega_{\text{MP}_n}\hat{c}_n^\dagger\hat{c}_n} \nonumber \\ 
    &+ i\left\lbrack\left(\overline{g}{\hat{b}}^{\dagger}+\sum_{n}{{\overline{g}}_n\hat{c}_n^{\dagger}} \right)\left({\hat{a}}_++{\hat{a}}_-^\dagger\right) \right. \nonumber \\
    &\quad \left. -\left(\overline{g}\hat{b} + \sum_{n}{{\overline{g}}_n{\hat{c}}_n}\right)\left({\hat{a}}_-+{\hat{a}}_+^\dagger\right)\right\rbrack\nonumber \\ 
    &+ D\left({\hat{a}}_-+{\hat{a}}_+^\dagger\right)\left({\hat{a}}_++{\hat{a}}_-^\dagger\right),
\end{align}
where $\hbar$ is the reduced Planck constant, $\hat{a}^\dagger$ ($\hat{a}$) in the first term is the creation (annihilation) operator for the cavity photons of CR-active ($\xi=+$) and CR-inactive ($\xi=-$) circularly polarized modes, the second term represents the energy of the collective CR excitation of the 2DEG, $\hat{b}^\dagger$ ($\hat{b}$) being its creation (annihilation) operator, MP modes of different orders $n=1, 3$ are introduced in the third term, where $\hat{c}_n^\dagger$ ($\hat{c}_n$) and $\omega_{\text{MP}_n}$ are the corresponding creation (annihilation) operator and resonance angular frequency, and the fourth term is the light--matter interaction term considering the coupling of CR and all MP modes with the single cavity mode. 

Here, ${\overline{g}}$ and ${\overline{g}}_n$ are the coupling strengths of CR and the $n$-th MP mode with the cavity mode, respectively. Using the zero-detuning coupling strengths, $g$ and $g_n$, which are independent of the external $B$, ${\overline{g}}$ and ${\overline{g}}_n$ can be written as 
\begin{align}
    \overline{g} &= \sqrt{\frac{\omega_{\text{c}}}{\omega_0}}g \\
    {\overline{g}}_{n} &= \sqrt{\frac{\omega_{\text{MP}_n}}{\omega_0}}g_{n}.
\end{align}
The coefficient $D$ of the diamagnetic term is
\begin{align}
    D = \frac{{\overline{g}}^{2}}{\omega_{\text{c}}} + \sum_{n}^{}\frac{{\overline{g}}_{n}^{2}}{\omega_{\text{MP}_n}} = \frac{g^{2} + \sum_{n}^{}g_{n}^{2}}{\omega_0}.
\end{align} 
The coupling strengths appear in the relative permittivity as
\begin{align}
    \varepsilon(\omega) &= \varepsilon_{\text{bg}} - \frac{\omega_{\text{pl}}^{2}}{\omega\left( \omega - \omega_{\text{c}} + \frac{i}{\tau} \right)} \\ &-\sum_n\frac{\omega_{\text{pl}_n}^{2}}{\omega\left( \omega - \omega_{\text{MP}_n} + \frac{i}{\tau_n} \right)}, \label{permittivity}
\end{align}
where the plasma frequencies of CR and $n$-th MP mode, $\omega_{\text{pl}}$ and $\omega_{\text{pl}_n}$, have relations with the $B$-independent coupling strengths as 
\begin{align}
    g^{2} &= \frac{\omega_{\text{pl}}^{2}d_{\text{QW}}}{\varepsilon_{\text{bg}}L_{\text{eff}}} \\
    g_{n}^{2} &= \frac{\omega_{\text{pl}_n}^{2}d_{\text{QW}}}{\varepsilon_{\text{bg}}L_{\text{eff}}}    
\end{align}
Here, $\varepsilon_{\text{bg}}=\varepsilon_\text{r}$ is the background relative permittivity, $\tau$ ($\tau_n$) is the lifetime of CR (MP$_n$), $d_{\text{QW}}$ is the thickness of the quantum well, and $L_{\text{eff}}$ is the effective cavity length. Using the above relative permittivity, the transmission spectra can be obtained through the transfer-matrix method. 

On the other hand, the polariton dispersion can be calculated from the $B$-dependent coupling strengths. The Heisenberg equations for the operators $\hat{a}_+$, $\hat{a}_-^{\dagger}$, $b$, and $c_n$ form a closed set describing the CR-active polariton modes, while those of $\hat{a}_-$, $\hat{a}_+^{\dagger}$, $b^{\dagger}$, and $c_n^{\dagger}$ describe the CR-inactive modes. The CR-active polariton annihilation operator is given by the Bogoliubov transformation, ${\hat{p}}_{+} = w{\hat{a}}_{+} + y{\hat{a}}_{-}^{\dagger} + x\hat{b} + \sum_{n}^{}{x_{n}{\hat{c}}_{n}}$. In the same manner as performed by Ciuti \textit{et al}.~\cite{ciutiQuantumVacuumProperties2005}, the polariton frequencies and the coefficients are determined by solving the eigenvalue problem of the following matrix:
\begin{align}
\begin{pmatrix}
\omega_0 + D & - D & i\overline{g} & i{\overline{g}}_{1} & \cdots \\
D & - \omega_0 - D & i\overline{g} & i{\overline{g}}_{1} & \cdots \\
 - i\overline{g} & i\overline{g} & \omega_{\text{c}} & 0 & \cdots \\
 - i{\overline{g}}_{1} & i{\overline{g}}_{1} & 0 & \omega_{\text{MP}_1} & \cdots \\
 \vdots & \vdots & \vdots & \vdots & \ddots \\
\end{pmatrix}.
\end{align}

\begin{figure}[t!]
\includegraphics[width=\columnwidth]{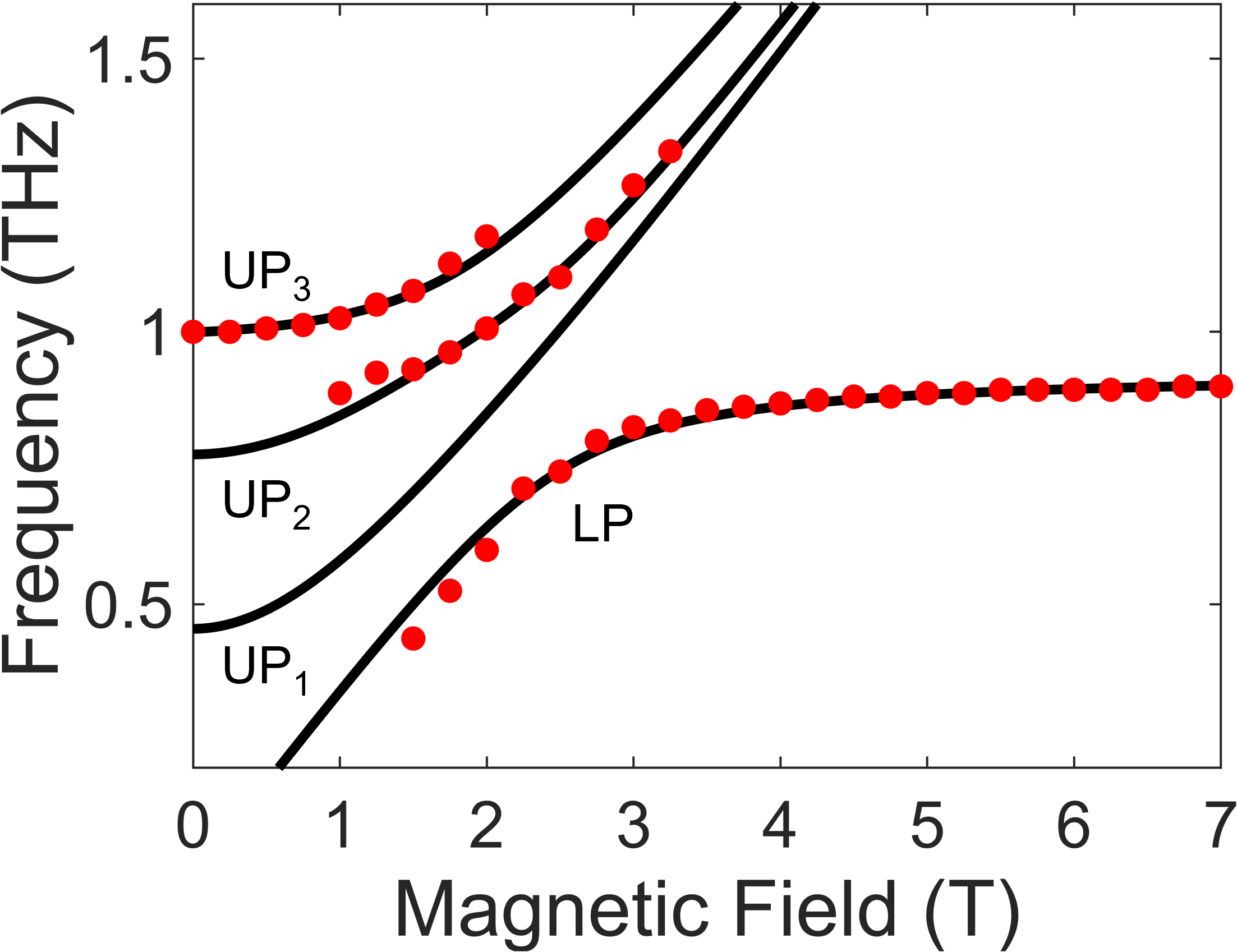}
\caption{Theoretical analysis of the multimode coupled system. Theoretical polariton frequencies based on the multimode Hopfield model. Red dots are extracted peak frequencies from experiment.}
\label{fig3}
\end{figure}

Theoretical transmission spectra for the multimode Landau polariton system, calculated using the transfer-matrix method with the relative permittivity $\varepsilon(\omega)$ derived in Eq.\,\eqref{permittivity}, are shown in Fig.\,2(b). The resonator structure is simplified to a Fabry--P\'erot cavity; details of the calculation setup are provided in the Supplementary Materials. The splitting in the UP branch is observed and captures the key feature of the experimental data. The discrepancy in linewidth is associated with the difference in cavity geometry, with the slots having a reduced $Q$-factor compared to an ideal Fabry--P\'erot cavity. From fitting to the experimental peak frequencies, we obtained the normalized coupling strengths of $g/\omega_0=0.18$ and $g_1/\omega_0=g_3/\omega_0=0.084$, confirming the USC of the CR and the SC of MP modes with the cavity mode. Figure\,3 shows the calculated polariton frequencies (black solid lines) overlaid with the peak frequencies extracted from the experiments (red dots). Theoretical predictions match the experimental results well. We note that the assumption $g_1=g_3$ is generally not valid but is enough to explain the experimental behavior. 

We further validated our experimental observation and theoretical model using three-dimensional finite-element simulations in COMSOL. 
Figure\,4 shows a color map of transmittance simulated with $m^*=0.070m_0$ and $\varepsilon_\text{r}=9.77$. Compared with the experimental transmittance in Fig.\,2(a), the UP branch is split into three parts, separated not only by the MP$_3$ mode but also by MP$_5$, resulting in UP$_4$. Such influence of the MP$_5$ mode could not be observed experimentally due to the weak magnitude. Incorporating the MP$_5$ contribution to the multimode Hopfield model, theoretical fits to the polariton frequencies are calculated, shown in black dashed lines in Fig.\,4; see Supplementary Materials for more detail. The obtained coupling strengths are $g/\omega_0=0.13$ and $g_1/\omega_0=g_3/\omega_0=g_5/\omega_0=0.062$. The slight deviation from the experimentally extracted values likely originates from uncertainties in the sample parameters and the strong effect of the MP$_5$ mode. Overall, the simulation is consistent with both the experimental and theoretical analysis of the multimode coupling. 

\begin{figure}[t!]
\includegraphics[width=\columnwidth]{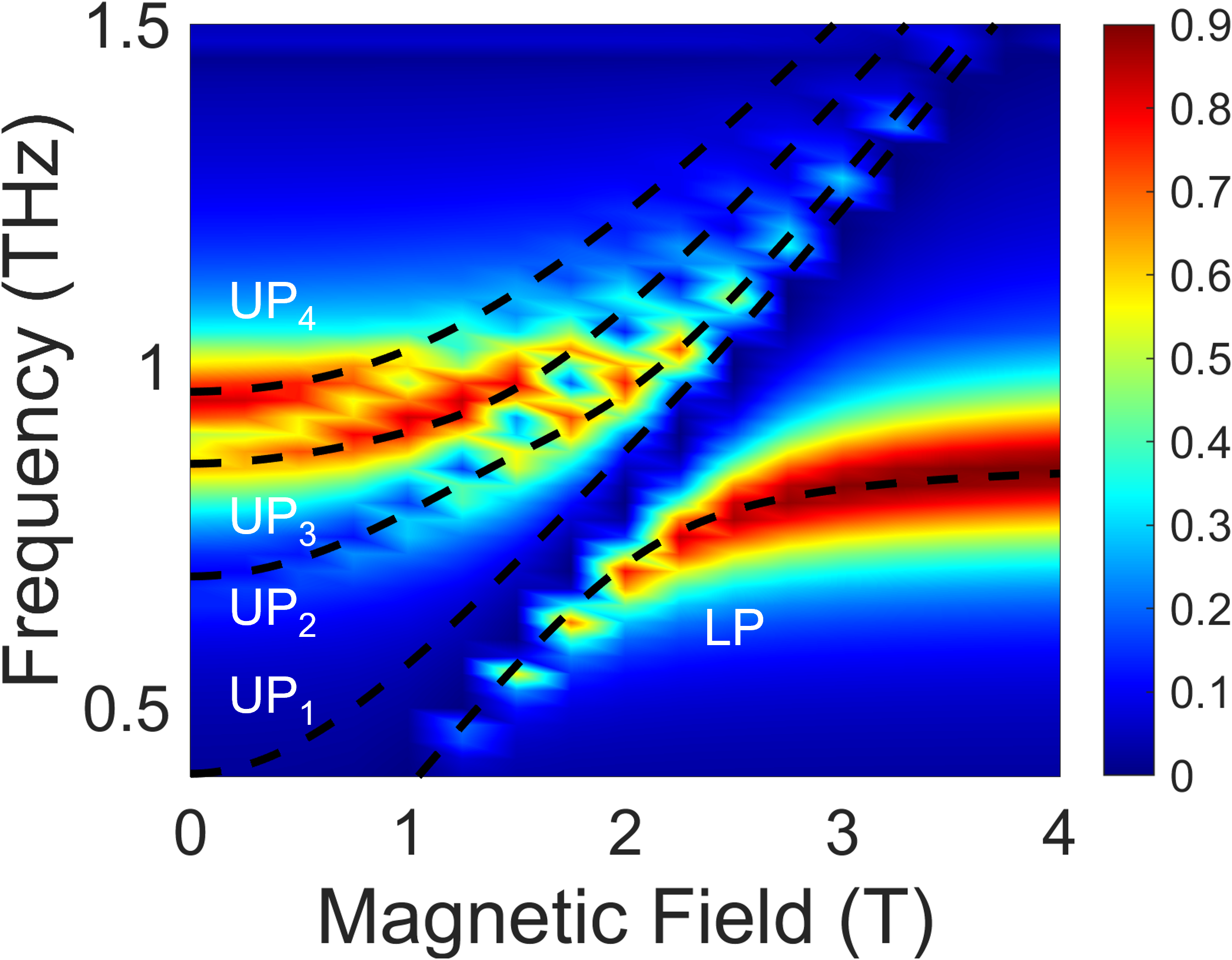}
\caption{Finite-element simulation of the multimode coupled system. Simulated color map of transmittance as a function of frequency and magnetic field using COMSOL. The black dashed lines are theoretical fits to the polariton frequencies based on the multimode Hopfield model. An additional upper polariton mode, denoted UP$_4$, appears due to the influence of the MP$_5$ mode.}
\label{fig4}
\end{figure}
    In conclusion, we experimentally explored the coupling between two distinct matter modes mediated by a cavity mode in an ultrastrongly coupled Landau polariton system. Owing to the small-mode-volume slots, cavity photons could interact with both the CR at $k=0$ and MP modes at finite $k$. Such coexistence of CR and MP excitations has been discussed in previous studies using 2DEG stripes and periodic metasurfaces~\cite{paravicini-baglianiGateMagneticField2017, mornhinwegModemultiplexingDeepstrongLightmatter2024, mornhinwegSculptingUltrastrongLight2024}. However, the multimode interaction demonstrated here could not be accessed due to the excitation geometry and frequency mismatch. In this work, the MP modes affected the Landau polariton through multimode hybridization and led to a splitting of the UP branch. From fitting, we obtained a normalized coupling strength of $g/\omega_0=0.18$ for CR, and $g_1/\omega_0=g_3/\omega_0=0.084$ for MP modes with an in-plane wave vector of $k=\pi/d$ and $k=3\pi/d$, showing the USC of CR and SC of MP modes to the cavity mode. The experimental results were well explained by the extended multimode Hopfield model and three-dimensional finite-element simulations. Our work establishes a foundation for controlling cavity-mediated nonlocal correlations between matter excitations in the USC regime, unlocking new possibilities for tailoring light--matter interactions. 
\\ \\
\textbf{Acknowledgments:}
S.R.E.\ acknowledges support from TOMODACHI--Dow Women's STEM Leadership and Research Program.  D.K.\ acknowledges Wen-Hua~Wu, Gustavo~Rodriguez~Barrios, and Shuying~Chen for technical assistant for experiments, and Fuyang~Tay for fruitful discussion. D.L. acknowledges UM6P for providing computational resources (HPC-TOUBKAL).

\noindent\textbf{Research funding:}
D.K.\ and J.K.\ acknowledge support from the U.S.\ Army Research Office (through Award No.\ W911NF2110157), the W.\ M.\ Keck Foundation (through Award No.\ 995764), the Gordon and Betty Moore Foundation (through Grant No.\ 11520), and the Robert A.\ Welch Foundation (through Grant No.\ C-1509). M.B.\ acknowledges support from the Research Foundation for Opto-Science and Technology and from the Japan Society for the Promotion of Science (JSPS) Grant No.\ JPJSJRP20221202 and KAKENHI Grant No.\ JP24K21526, JP25K00012, JP25K01691, and JP25K01694.

\noindent\textbf{Author contributions:}
J.K.\ supervised the project. S.R.E., D.K., and A.C.-M.\ performed all THz measurements. S.R.E.\ and D.K.\ analyzed the experimental data under the guidance of J.K. D.K.\ built the THz magnetospectroscopy setup. S.L.\ grew the quantum well under the guidance of M.J.M. G.L., S.K., and D.K.\ designed and fabricated the slot cavities under the guidance of M.S. M.B.\ developed the theoretical model. D.L.\ performed numerical simulations and analyzed the experimental data. S.R.E., D.K., M.B., and J.K.\ wrote the manuscript. All authors discussed the results and commented on the manuscript. All authors have accepted responsibility for the entire content of this manuscript and approved its submission. 

\noindent\textbf{Conflict of interest:}
Authors state no conflicts of interest.

\noindent\textbf{Data availability:}
The datasets generated and/or analyzed during the current study are available from the corresponding author upon reasonable request.

\bibliography{ref3,Dasom} 
\clearpage
\section*{Supplementary Materials}
    \subsection*{S1 Details of the theoretical calculation setup} 
 
Figure S1 shows the simplified structure of the multimode Landau polariton system adopted in the theoretical calculations. Instead of slot cavities, we simply consider a Fabry--P\'erot cavity embedding a 2DEG under a magnetic field. The material parameters are taken from the experiment, with the magnetoplasmon lifetime at 1.2 ps. The GaAs layers have a permittivity of 3.6 and a thickness of $d_\text{GaAs}=22.35$\,$\upmu$m. The permittivity of the gold layers is calculated using the Drude model. The bare cavity frequency is adjusted to match the experimental peak position at 7\,T. The calculated effective cavity length is 84.2\,$\upmu$m. The transmission spectra and the polariton dispersion are obtained with the same parameters. 

For fitting to the simulated spectra, $d_\text{GaAs}$ is set to 23.85\,$\upmu$m, while the other parameters follow those used in the COMSOL simulation. The bare cavity frequency is set to 0.870 THz. The calculated effective cavity length is 58.0\,$\upmu$m. 

\renewcommand{\thefigure}{S\arabic{figure}}
\setcounter{figure}{0} 
\begin{figure}[h]
\includegraphics[width=\columnwidth]{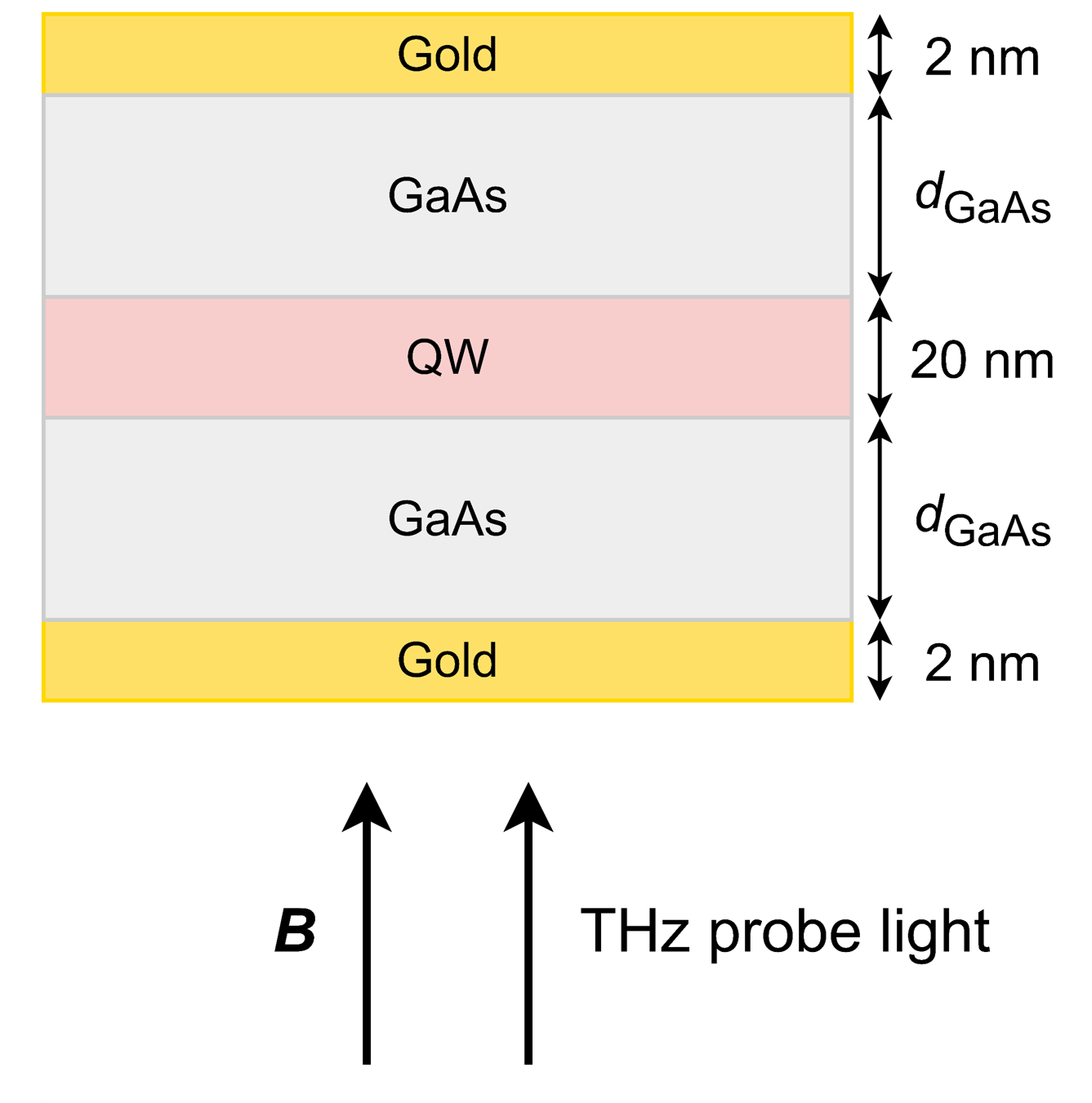}
\caption{Schematic of the multimode Landau polariton system used for the theoretical calculations.}
\label{figs1}
\end{figure}
\clearpage
\onecolumn
\end{document}